\documentclass[12pt]{article}
\usepackage{amsmath}
\begin{document}

\pagestyle{myheadings}
\markright{F. Wilczek \hfill \emph{Some Basic Aspects of 
 Fractional Quantum Numbers}\qquad}
\thispagestyle{empty}

\title{Some Basic Aspects of\\
 Fractional Quantum Numbers\footnote{Commentary for the Volume,
``Collected Works of J. Robert Schrieffer''}}
\author{Frank Wilczek\\
\small\it  Center for Theoretical Phyics\\
\small\it Massachusetts Institute of Technology\\
\small\it Cambridge, MA 02139-4307\\
\small MIT-CTP-3275\qquad cond-mat/0206122}
\date{}
\maketitle

\begin{abstract}\noindent
I review  why  and how physical states with fractional quantum numbers can occur, 
emphasizing basic mechanisms in simple contexts.   
The general mechanism of charge fractionalization is the passage from states created by local action of fields to states
having a topological character, which permits mixing between local and topological charges.  The primeval case of charge
fractionalization for domain walls, in polyacetylene and more generally, can be demonstrated convincingly using  Schrieffer's intuitive counting
argument, and derived formally from analysis of zero modes and vacuum polarization.   An important generalization realizes
chiral fermions on high-dimensional domain walls, in particular for liquid He3 in the A phase.    In two spatial dimensions,
fractionalization of angular momentum and quantum statistics occurs, for reasons that can be elucidated both 
abstractly, and specifically in the context of the quantum Hall effect. 
\end{abstract}

%
%\pagebreak

\newpage

Quantization of charge is a very basic feature of our picture of the physical world.  
The explanation of how matter can be built up from a few types of indivisible building-blocks, each occurring in
vast numbers of identical copies, is a major triumph of local quantum field theory.  In many ways, it forms the
centerpiece of twentieth century physics. 

Therefore the discovery of physical circumstances in which the unit of charge can be fractionated, its quanta dequantized, came as a shock
to most physicists.   It is remarkable that this fundamental discovery emerged neither from recondite theoretical speculation, nor from
experiments at the high-energy frontier, but rather from analysis of very concrete, superficially mundane (even messy) polymers \cite{JR, SSH}.   In
the process of coming to terms with charge fractionalization, we've been led to a deeper understanding of the logic of charge
quantization itself.  We have also been led to discover a whole world of related, previously unexpected phenomena.  Exploration
of this concept-world is far from complete, but already it has proved richly rewarding, and fed back into the description of
additional real world phenomena.  

Bob Schrieffer's contributions in this field, partially represented in the papers that follow, 
started early and have run deep and wide.  
In this introduction I've attempted to distill the core theoretical concepts to their simplest, 
most general meaningful form, and put them in a
broader perspective.  Due to limitations of time, space and (my) competence, serious analysis of 
particular materials and their experimental
phenomenology, which figures very prominently in Schrieffer's papers, will not be featured here.

\section{The Secret of Fractional Charge}

To begin, let us take a rough definition of
charge to be any discrete, additive, effectively conserved quantity, and let us accept the conventional story of charge
quantization  as background.   A more discriminating discussion of different varieties of charge, and of the origin of quantization,
follows shortly below.  

The conventional story of charge quantization consists of three essential points: some deep theory gives us a universal unit for the charges to
be associated with fields; observed particles are created by these fields, acting locally; the charges of particles as observed are related by
universal renormalization factors to the charges of the fields that create them.    The last two points are closely linked.  Indeed, conservation
of charge implies that the state produced by a local field excitation carries the charge of the field.   Thus renormalization of charge reflects
modification of the means to measure it, rather than of properties of the carriers.  This is the physical content of Ward's identity, leading to the relation
$e_{\rm ren.} = (Z_3)^{\frac{1}{2}} e_{\rm bare}$ between renormalized and bare charge in electrodynamics, wherein only  wave-function
renormalization of the photon appears.   

This reasoning, however, does not apply to states that cannot be produced by local action of quantum fields, which often occur.   Such states
may, for example, be associated with topologically non-trivial rearrangements of the conditions at infinity.   Simple, important examples are
domain walls between two degenerate phases in 1 spatial dimension systems and flux tubes in 2 spatial dimension systems.  These states are
often associated additive quantum numbers, also called topological charges.   For example, the flux itself is an additive quantum number
classifying flux tubes, given in terms of gauge potentials at spatial infinity by $\oint d\theta  A_\theta$.   

With two underlying charges, the general
relation between renormalized and bare charge becomes
\begin{equation}
q_{\rm ren.} = \epsilon_1 q^{(1)}_{\rm bare} + \epsilon_2 q^{(2)}_{\rm bare} .
\end{equation}
Given this form, quantization of $q^{(1)}_{\rm bare}$ and $q^{(2)}_{\rm bare}$ in integers does not imply that renormalized charges are
rationally related.  In particular, suppose that the first charge is associated with local fields, while the second is topological.  Then ratio of
renormalized charges for a general state, and the state of minimal charge produced by local operations is a conventionally normalized
charge.  In the absence of topology, it would be simply $q^{(1)}$, an integer.  Here it becomes
\begin{equation}
q_{\rm normalized} = q^{(1)} + \frac{\epsilon_2}{\epsilon_1} q^{(2)} .
\end{equation}
The ratio $\frac{\epsilon_2}{\epsilon_1}$ is a dynamical quantity that, roughly speaking, measures the
induced charge associated with a unit of topological structure.  In general, it need not be integral or even
rational.  This is the general mechanism whereby fractional charges arise.  It is the secret of fractional
charge.  

An important special case arises when the topological charge is discrete, associated with a finite additive group $Z_n$.  
Then the renormalized
charge spectrum for $q^{(2)} = n$ must be the same as that for $q^{(2)} =0$, since a topological charge $n$ configuration, being
topologically trivial, can be produced by local operations.  So we have the restriction 
\begin{equation}
\frac{\epsilon_2}{\epsilon_1} = \frac{p}{n}
\end{equation}
with an integer $p$.   Then the fractional parts of the normalized charges are always multiples of $\frac{1}{n}$.  

The primeval case of 1 dimensional domain walls, which we are about to discuss in depth, requires a special comment.  A domain
wall of the type $A\rightarrow B$, going from the $A$ to the $B$ ground state, can only be followed (always reading left to right) by an
anti-domain wall of the type $B\rightarrow A$, and {\it vice versa}; one cannot have two adjacent walls of the same type.  So one does not
have, for domain wall number, quite the usual physics of an additive quantum number, with free superposition of states.   
However, the underlying, spontaneously broken
$Z_2$ symmetry that relates $A$ to $B$ also relates domain walls to anti-domain walls.  Assuming that this symmetry commutes with the
charge of interest, the charge spectra for domain wall and the anti-domain wall must agree.   (This assumption is valid in the case at hand;
indeed, the charges whose values are of most interest generally are those associated with unbroken symmetries.)  At the same time, the
spectrum of total charge for domain wall plus anti-domain wall, a configuration that can be produced by local operations, 
must reduce to that
for vacuum.  So we have
$2\frac{\epsilon_2}{\epsilon_1} = {\rm integer}$, and we find (at worst) half-integer normalized charges, just as if the domain wall charge
were itself a proper $Z_2$ charge.

\section{Polyacetylene and the\\ Schrieffer Counting Argument}

For our purposes, polyacetylene in its ground state can be idealized as an infinite chain molecule with alternating single and
double bonds.   This valence structure is reflected, physically, in the spacing of neighboring carbon nuclei: those linked by
double bonds are held closer than those linked by single bonds.  Choosing some particular nucleus to be the origin, and
moving from there to the right, there are two alternative ground states of equal energy, schematically
\begin{eqnarray}
\cdots 121212121212 \cdots ({\rm A})\cr
\cdots 212121212121 \cdots ({\rm B})
\end{eqnarray}
Now consider the defect obtained by removing a bond at the fourth link, in the form
\begin{equation}
\cdots 121112121212 \cdots 
\end{equation}
By shifting bonds down between the tenth and fifth links we arrive at
\begin{equation}
\cdots 121121212112 \cdots , 
\end{equation}
which displays the original defect as two more elementary ones.   Indeed, the elementary defect
\begin{equation}
\cdots 12112121 \cdots , 
\end{equation}
if continued without further disruption of the order,  is a minimal domain wall interpolating between 
ground state A on the left and ground state B on the right.  

The fact that by removing {\it one\/} bond we produce {\it two\/} domain walls strongly suggests that each domain wall is
half a bond short.  If bonds were electrons, then each wall would fractional charge $\frac{e}{2}$, and spin $\pm
\frac{1}{4}$.   In reality bonds represent pairs of electrons with opposite spin, and so we don't get fractional charge.  But we still
do find something quite unusual: a domain wall acquires charge $e$, with spin 0.  Charge and spin, which normally occur together, have
been separated!    

This brilliant argument, both lucid and suggestive of generalizations, is known as the Schrieffer counting argument.  
In it, the secret of fractional charge is reduced to barest bones.  

A simple generalization, which of course did not escape Schrieffer \cite{SS}, is to consider more elaborate bonding possibilities, for
example
\begin{equation}
\cdots 112112112112 \cdots
\end{equation}
Here removing a bond leads to 
\begin{equation}
\cdots 111112112112 \cdots
\end{equation}
 which is re-arranged to 
\begin{equation}
\cdots 111211121112 \cdots ,
\end{equation}
containing {\it three\/} elementary defects.   Clearly true fractions, involving one-third integer normalized electric charges, are now
unavoidable.

\section{Field Theory Models of Fractional Charge}

While the Schrieffer counting argument is correct and utterly convincing, it's important and fruitful to see how its results are
realized formally, in quantum field theory.

First we must set up the field theory description of polyacetylene.  
Here I will very terse, since the accompanying paper of Jackiw and Schrieffer sets out this problem in detail \cite{JS}.  We consider a half-filled band
in one dimension.  With uniform lattice spacing $a$, the fermi surface consists of the two points $k_\pm = \pm \pi/2a$.  We can parametrize
the modes near the surface using a linear approximation to the energy-momentum dispersion relation; then near these two points we have
respectively right- and left-movers with velocities $\pm |\frac{\partial \epsilon}{\partial k}|$.  Measuring velocity in this unit, and
restricting ourselves to these modes, we can write the free theory in pseudo-relativistic form.  (But note that in these considerations,
physical spin is regarded only as an inert, internal degree of freedom.)   It is convenient here to use the Dirac matrices
\begin{eqnarray}
\gamma^0 ~=~& \left( \begin{matrix} 0&1 \\ \ 1&0 \end{matrix} \right) \cr \cr
\gamma^1 ~=~& \left( \begin{matrix} 0&{-1} \\ 1&0 \end{matrix} \right) \cr \cr
\gamma^{\chi} ~=~&\left( \begin{matrix} 1&0 \\ 0&-1 \end{matrix} \right) ,
\end{eqnarray}
where $\gamma^{\chi} \equiv \gamma^0 \gamma^1$ is used to construct the chirality projectors $\frac{1 \pm \gamma^\chi}{2}$.  In the
kinetic energy 
\begin{equation}
L_{\rm kinetic} ~=~ \bar \psi (i \gamma \cdot \partial) \psi
\end{equation}
the right- and left-movers $\frac{1\pm \gamma^{\chi}}{2} \psi$ do not communicate with one another.   However scattering on the optical phonon
mode
$\phi$, with momentum $\pi /a$, allows electrons to switch from one side of the fermi surface to the other.  This is represented by the local Yukawa
interaction
\begin{equation}
\Delta L (x, t) = g \phi (x,t ) \bar \psi (x, t)  \psi (x,t) . 
\end{equation}
One also has kinetic terms for $\phi$ and a potential $V(\phi)$ that begins at quadratic order.  The wave velocity for $\phi$ of course need not
match the fermi velocity, so there is a violation of our pseudo-relativistic symmetry, but it plays no role in the following.  

This field theory description does not
yet quite correspond to the picture of polyacetylene sketched in the previous section, because the breaking of translational symmetry (from
$x\rightarrow x+a$ to $x\rightarrow x+2a$) has not appeared.   We need not change the equations, however, we need only draw out their
implications.   Since our optical phonon field $\phi$ moves neighboring nuclei in opposite directions, a condensation $\langle \phi
\rangle = \pm v \neq 0$ breaks translational symmetry in the appropriate way.  We might expect this symmetry breaking
phonon condensation to be favorable, at half filling, because it opens up a gap at the fermi surface, and lowers the energy of
occupied modes near the top of the band.   Since these modes have been retained in the effective field theory, the instability
should be reflected in that theory.   Calculation bears out these intuitions.  The classical potential $V(\phi )$ is
subject to quantum corrections, which alter it qualitatively.   Upon calculation of a simple one-loop vacuum polarization graph,
one finds a correction 
\begin{equation}
\Delta V(\phi) = \frac{g^2}{\pi} \phi^2 \ln (\phi^2 / \mu^2),
\end{equation}
where $\mu$ is an ultraviolet cutoff.   (This cutoff appears because the assumed Yukawa interaction $g\phi \bar \psi \psi$ 
is an appropriate description of physics
only near the fermi surface.  A more sophisticated treatment would use the language of the renormalization group here.)  For small $\phi$ this
correction always dominates the classical $\phi^2$.  So it is always advantageous for $\phi$ to
condense, no matter how small is $g$.   Indeed, one finds the classic ``BCS type'' dependence
\begin{equation}
\langle \phi \rangle ^2  = \mu^2 e^{-\frac {m^2 \pi}{g^2}}
\end{equation}
at weak coupling.   

This elegant example of dynamical symmetry breaking was first discussed by Peierls \cite{P}, who used a rather
different language.  It was introduced into relativistic quantum field theory in the seminal paper of Coleman and E. Weinberg \cite{CW}.  In
four space-time dimensions the correction term goes as $\Delta V(\phi) \propto g^4 \phi^4 \ln \phi^2$, and it dominates at small
$\phi$ only if the classical mass term ($\propto \phi^2$) is anomalously small.  

\subsection{Zero Modes}

The symmetry breaking $\langle \phi \rangle = \pm v$ induces, through the Yukawa coupling, an effective mass term for the
fermion $\psi$, which of course can be interpreted in the language of condensed matter physics as the opening of a gap.   The
choice of sign, of course, distinguishes between two degenerate ground states that have identical physical
properties, since they can be related by the symmetry 
\begin{eqnarray}
\phi \rightarrow & -\phi \cr
\psi_L \rightarrow &  -\psi_L 
\end{eqnarray}
With this interpretation, we see that a domain wall interpolating between $\langle \phi (\pm \infty ) \rangle =  \pm v$,
necessarily has a region where the mass vanishes, and we might expect it to be favorable for fermions to bind there.    What is
remarkable, is that there is always a solution of zero energy -- a mid-gap state -- localized on the wall.  Indeed, in the background
$\phi(x) = f(x) $ the Dirac equation for zero energy is simply
\begin{eqnarray}
i \partial_x \psi_1 =& g f  \psi_2 \cr
i \partial_x \psi_2 =& g f  \psi_1
\end{eqnarray}
with the normalizable solution
\begin{eqnarray}
\psi_1(x) =& \exp (-g \int_0^x dy f(y) ) \cr
\psi_2(x) =& -i\psi_1(x) . 
\end{eqnarray}
Note that the domain wall asymptotics for $\langle \phi(x) \rangle$ allows the exponential to die in both directions.  

It is not difficult to show, using charge
conjugation symmetry (which is not violated by the background field!), that half the spectral weight of this mode arise from modes that are
above the gap, and half from modes that are below the gap, with respect to the homogeneous ground state.  

When we quantize the fermion field, we must decide whether or not to occupy the zero-energy mode.   If we occupy it, then
we will have occupied half a mode that was unoccupied in the homogeneous ground state, and we will have a state of fermion number
$\frac{1}{2}$.  If we do not occupy it, we will have the charge conjugate state, with fermion number $-\frac{1}{2}$.   It is
wonderful how this delicate mechanism, discovered by Jackiw and Rebbi \cite{JR}, harmonizes with the Schrieffer counting argument.

\subsubsection{Zero Modes on Domain Walls}

An abstract generalization of this set-up, with relativistic kinematics, is very simple, yet it has proved quite important.  Consider massless, relativistic
fermions in an odd number $2n+1$ of Euclidean dimensions, interacting with a scalar field $\phi$ according to $L_{\rm int.} = g \phi \bar \psi \psi$
as before.   Let $\langle \phi(z)\rangle = h(z)$ implement a domain wall, with $h(\pm \infty) = \pm v$.  Then off the wall the fermion acquires
mass$^2$
$m^2 = g^2 v^2$.  But, guided by previous experience, we might expect low-energy modes localized on the wall.  
Here we must look for solutions of the
$2n+1$ dimensional Dirac equation that {\it also\/} satisfy the $2n$ dimensional Dirac equation.   
With the factorized form $\psi(x, z) = f(z) s(x)$, where $f$ is a c-number and $s$ a spinor satisfying the $2n$-dimensional Dirac equation, we must
require  
\begin{equation}
\gamma^{2n+1} \frac{\partial}{\partial z} f(z)s ~=~  - g h(z) s .
\end{equation}
For $s_{\pm}$ an eigenspinor of $\gamma^{2n+1}$ with eigenvalue $\pm 1$, this leads to 
\begin{equation}
f_{\pm} (z) ~\propto~ e^{- \int_0^z dy (\pm g h(y) )}.
\end{equation}
Only the upper sign produces a normalizable solution.   Thus only a particular {\it chirality\/} of $2n$-dimensional spinor appears.   This mechanism
has been used to produce chiral quark fields for numerical work in QCD \cite{K}, avoiding the notorious doubling problem, and it has appeared in many
speculations about the origin of chirality in Nature, as it appears in the Standard Model of particle physics.

A very much more intricate example of chiral zero modes on domain walls, in the context of superfluid
He3 in the A phase, is analyzed in the accompanying paper of Ho, Fulco, Schrieffer and Wilczek
\cite{HFSW}.  (Note the date!)  A very beautiful spontaneous flow effect is predicted in that paper,
deeply analogous (I believe) to the persistent flow of edge currents in the quantum Hall effect.   I'm not
aware that this particular experiment, which is surely not easy, was ever carried through.  But,
especially in view of the advent of exquisitely controlled condensates of cold atoms,  I'm confident that
we haven't yet heard the last word on this subject, neither theoretically nor experimentally.  

\subsection{Vacuum Polarization and Induced Currents}

To round out the discussion, let us briefly consider a natural generalization of the previous model, to include two scalar fields $\phi_1, \phi_2$ and
an interaction of the form 
\begin{equation}
L_{\rm int.} ~=~ g_1 \phi_1 \bar \psi \psi + g_2 \phi_2 \bar \psi \gamma^\chi \psi .
\end{equation} 
Gradients in the fields $\phi_1, \phi_2$ will induce non-trivial expectation values of the number current 
$j^\mu \equiv \bar \psi \gamma^\mu \psi$ in the local ground state.   In the neighborhood of space-time points $x$ where the local value of the
effective mass$^2$, that is
$g_1^2 \phi_1^2 + g_2^2 \phi_2^2$, does not vanish, one can expand the current in powers of the field gradients over the effective mass.  To first
order, one finds 
\begin{align}
\langle j_\mu \rangle &= \frac{1}{2\pi} \frac {g_1 g_2 ( \phi_1 \partial_\mu \phi_2 - \phi_2
\partial_\mu \phi_1)}{g_1^2 \phi_1^2 + g_2^2
\phi_2^2}\nonumber \\
&= \frac{1}{2\pi} \partial_\mu \theta\\ 
\intertext{where}
\theta &\equiv \arctan \frac{g_2 \phi_2}{g_1 \phi_1} .
\end{align}

We can imagine building up a topologically non-trivial field configuration adiabatically, by slow variation of the $\phi$s.   As long as the
effective mass does not vanish, by stretching out this evolution we can justify neglect of the higher-order terms.  Flow of current at infinity is not
forbidden.  Indeed it is forced, for at the end of the process we find the accumulated charge 
\begin{equation}
Q = \int j^0 = \frac{1}{2\pi} (\theta (\infty) - \theta(-\infty)) 
\end{equation} 
on the soliton.   This, of course, can be fractional, or even irrational.  In appropriate models, it justifies Schrieffer's generalized counting argument
\cite{GW}. 

Our previous model, leading to charge $\frac {1}{2}$, can be reached as a singular limit.  One considers configurations where
$\phi_1$ changes sign with $\phi_2$ fixed, and then takes $g_2 \rightarrow 0$.  
This gives $\Delta \theta = \pm \pi$, and hence $Q=\pm \frac{1}{2}$, depending on which side the limit is approached from.

\section{Varieties of Charge}

In physics, useful charges come in several varieties -- and it seems that all of them figure prominently in the story of
fractional charge.   Having analyzed specific models of charge fractionalization, let us pause for a quick survey of the varieties of charge. 
This will both provide an opportunity to review foundational understanding of charge quantization, and set the stage for more intricate
examples or fractionalization to come.  

Deep understanding of the
issues around charge quantization can only be achieved in the context of quantum field theory.  Even the prior fact that there are many
entities with rigorously identical properties, for example many identical electrons, can only be understood in a satisfactory way at this level.   

\subsection{Bookkeeping Charges}

The simplest charges, conceptually, are based on counting.  They
encode strict, or approximate, conservation laws if the numbers thus calculated before and
after all possible, or an appropriate class of, reactions are equal.    Examples of useful charges based on counting are electric charge,
baryon number, lepton number, and in chemistry 90+ laws expressing the separate conservation of number of atoms of each element.

Using operators $\phi_j$ to destroy,  and their conjugates $\phi_j^+$ to create, particles of type $j$ with charge $q_j$, a strict
conservation law is encoded in the statement that interaction terms 
\begin{equation} 
\Delta L_{\rm int.} \sim \kappa \prod_m  \phi_{j_m}^{k_{j_m}} \prod_n  (\phi^+)_{j_n}^{k_{j_n}}
\end{equation}
which fail to satisfy 
\begin{equation}
\sum_m k_{j_m} q_{j_m} = \sum_n k_{j_n} q_{j_n}
\end{equation}
do not occur.  (In the first expression, already awkward enough, all derivatives and spin indices have
been suppressed.)  Alternatively, the Lagrangian is invariant under the abelian symmetry transformation 
\begin{equation}
\phi_j \rightarrow  e^{i \lambda q_j} \phi_j .
\end{equation}
An approximate
conservation law arises if such terms occur only with small coefficients.  One can also have discrete conservation laws, where the
equality is replaced by congruence modulo some integer.

In all practical cases effective Lagrangians are polynomials of small degree in a finite number of fields.   In that context, conservation
laws of the above type, that forbid some subclass of terms, can always be formulated, without loss of generality, 
using integer values of
the $q_j$.  It will be usually appear simple and natural to do so.   In a sense, then, quantization of charge is automatic.  
More precisely,
it is a consequence of the applicability of local quantum field theory at weak coupling, which is what brought us to this class of
effective Lagrangians.  

Of course, the fact that we {\it can\/} always get away with integers does not mean that we {\it must\/} do so.   For example,
suppose we have a situation where there are two applicable conservation laws, with integer charges $q^{(1)}_j, q^{(2)}_j$ for particles
of type $j$.  If I define the master-charge 
\begin{equation}
Q_j \equiv q^{(1)}_j + w q^{(2)}_j , 
\end{equation}
with $w$ irrational, then conservation of $Q$ encodes both of the prior conservation laws simultaneously.  This semi-trivial
trick touches close to the heart of the fractional charge phenomenon, as exposed above.  

\subsubsection{Gauge Charges; Nonabelian Symmetry}

Substantial physical issues, that are definitely not matters of convention, arise for conserved quantities that have independent
dynamical significance.   The prime example is electric charge, to which the electromagnetic field is sensitive.  

Empirically, the   
electric charges of electrons and protons are known to be equal and
opposite to within a part in $10^{-21}$.   
Their cancellation occurs with 
fantastic accuracy despite the fact that the protons and electrons are 
very different types of particles, and in particular despite the fact that
the proton is composite and is subject to the strong interaction.   
More generally, the accurate neutrality of all unionized atoms, not only hydrogen, can be tested with sensitive atomic
beam experiments, and has never been found to fail.  

Neither pure quantum electrodynamics nor its embedding into the Standard Model of matter explains why electrons and protons
carry commensurate charges, though of course both theories can accommodate this fact.   Specifically, either theory would retain its
intellectual integrity if the photon coupled to a modified charge 
\begin{equation}
\tilde Q = Q + \epsilon (B-L) ,
\end{equation}
where $B$ is baryon number, $L$ is lepton number, and $\epsilon$  a numerical coefficient, instead of to the conventional charge
$Q$.   If $\epsilon$ is taken small enough, the modified theories will even continue to agree with experiment.   

To produce a mandatory unit of charge, that cannot be varied by small amounts from particle to particle (or field to field), we must embed
the abelian counting symmetry into a simple, nonabelian group.   Unified gauge theories based on the gauge groups $SU(5)$ or $SO(10)$
accomplish this; moreover, they account nicely for the full spectrum of $SU(3)\times SU(2) \times U(1)$ quantum numbers for the
particles observed in Nature \cite{GG}.   This represents, at present, our best understanding of the origin of charge quantization.   It indirectly
incorporates Dirac's idea \cite{D} that the existence of magnetic monopoles would force the quantization of charge, since these theories
contain magnetic monopoles as regular solutions of the field equations \cite{'t, Po}.

\subsection{Topological Charges}

Bookkeeping charges, as described above, reside directly in quantum fields, and from there come to characterize the small-amplitude
excitations of these fields, that is the corresponding particles.  These particles are, at the level of the effective field theory,
point-like.   In addition to these objects, the theory may may contain collective excitations with a useful degree of stability, which then
become significant, identifiable objects in their own right.     These are usually associated with topological properties of the fields, and are
generically called solitons.   Of course, at the next level of description, solitons themselves can be regarded as primary ingredients in an
effective theory.

Solitons fall into two broad classes, boundary solitons and texture solitons.    Boundary solitons are associated with non-trivial structure at
spatial infinity.   A simple example is domain walls in polyacetylene, as discussed above.   Texture solitons are associated with non-trivial
mapping of space as a whole into the target field configuration space, with trivial structure at infinity.   A simple example is a phason in 1
space dimension, as covered implicitly above (take $\phi_1^2 + \phi_2^2 = {\rm constant}$).  
There the target space for the field $\theta$ is a circle, and a field configuration that starts with $\theta= 0$ on the far left and winds
continuously to $\theta = 2\pi$ on the far right has non-trivial topology as a mapping over all space, though none at the boundaries.   
Skyrmions \cite{S, R} provide a higher-dimensional generalization of this type.   Texture solitons can be produced by local operations, but generally not
by means of a finite number of field operations (creation and destruction operators) so their topological quantum numbers can also appear in
fractionalization formulae.

\subsection{Space-Time and Identity Charges}

Each of the charges we have discussed so far can be considered as a label for representations of some symmetry group.  This is
obvious for bookkeeping  charges, which label representations of phase groups; it is also true for topological charges, which label
representations of homotopy groups.   There are also symmetry groups associated with space-time transformations, specifically
rotations, and with interchange of identical particles.   And there are corresponding quantum numbers.  For rotations this is, of
course, spin.  For identity it is fermi versus bose character -- an additive, $Z_2$ quantum number.  These quantum numbers are quite
familiar to all physicists.  Less familiar, and perhaps unsettling on first hearing, is the idea that they can be dequantized.  Let's focus on that
now.

\subsubsection{Space-Time Charges}

In three space dimensions, rotations generate the nonabelian group $SO(3)$.  The quantization of spin, in integer units, follows from
this.  Actually, not quite -- that would prove too much, since we know there are particles with half-integer spin.    The point is that
quantum mechanics only requires that symmetry groups are implemented ``up to a phase'', or, in the jargon, projectively.  If the
unitary transformation associated with a symmetry generator $g$ is $U(g)$, then we need only have
\begin{equation}
U(g_1) U(g_2) = \eta (g_1, g_2) U(g_1 g_2) , 
\end{equation}
where $\eta (g_1, g_2)$ is a phase factor, since observables, based on inner products, will not depend on $\eta$.   It turns out that
projective representations of $SO(3)$ correspond to ordinary representations of $SU(2)$, so one still has quantization, but in
half-integer units.  

In two space dimensions the group is $SO(2)$.  We can parametrize its elements, of course, in terms of an angle $\theta$, and its
irreducible representations by the assignments $U(\theta ) = e^{is\theta}$, for $0\leq \theta < 2\pi$.   These are ordinary
representations only if $s$ is an integer; but they are perfectly good projective representations for any value of $s$.   Thus in two
space dimensions angular momentum is dequantized.

\subsubsection{Identity Charges}

Among all quantum-mechanical groups, perhaps the most profound is the symmetry group associated with interchange
of identical particles.  For the existence of this symmetry group, manifested in the existence of quantum statistics and
associated exchange phenomena, permits us to reduce drastically the number of independent entities we need to describe
matter.   

We teach undergraduates that quantum statistics supplies symmetry
conditions on the wave function for several identical particles: the
wave function for bosons must not change if we interchange the coordinates
of two of the bosons, while the wave function for fermions must be
multiplied by
-1 if we exchange the coordinates of two of the fermions.  If the interchange
of two particles is to be accompanied by a fixed phase factor $e^{i\theta}$, it would seem that this factor had better be $\pm
1$, since iterating the exchange must give back the original wave
function.  Nevertheless we can make sense of the notion of fractional statistics; but
to do so we must go back to basics \cite{LM}.  

In quantum mechanics we are required
to compute the amplitude for one configuration to evolve into another over 
the course of time.  Following Feynman, this is done by adding together the
amplitudes for all possible trajectories (path integral).  Of course the
essential dynamical 
question is: how are we to weight the different paths?  
Usually, we take
guidance from classical mechanics.  To quantize a classical system with
Lagrangian $L$ we integrate over all trajectories weighted by their
classical action $e^{i\int L dt}$.  However, essentially 
new possibilities arise when the space of trajectories falls into
disconnected pieces.  Classical physics gives us no guidance as to 
how to assign relative
weights to the
different disconnected pieces of trajectory space.  
For the classical equations of motion are
the result of comparing infinitesimally different paths, and 
in principle supply no
means to compare paths that cannot be bridged by a succession of
infinitesimal variations.  

The space of trajectories of identical particles, 
relevant to the question of quantum statistics,
does fall into
disconnected pieces.  Suppose, for example,
that we wish to construct the amplitude to
have particles at positions $x_1, x_2, ... $ at time $t_0$ and again at
time $t_1$.  The total amplitude gets contributions not only 
from trajectories such
that the particle originally at $x_1$ winds up at $x_1$, but also from
trajectories where this particle winds up at some other $x_k$, and its
place is taken up by a particle that started from some other position.  All
permutations of identity between the particles in the initial and final
configurations, are possible.  Clearly, trajectories that result in 
different permutations cannot be continuously deformed into one another. 
Thus we have the situation mentioned above, that the space of trajectories
falls into disconnected pieces.  

Although the classical limit cannot guide us in the choice of weights,
there
is an important consistency condition from quantum mechanics itself that
severely limits the possibilities.  We must respect the rule, that if
we follow a trajectory ${\alpha}_{01}$ from $t_0$ to $t_1$ by a trajectory 
$\alpha_{12}$ from $t_1$ to $t_2$, then the amplitude assigned to the
combined trajectory $\alpha_{02}$ should be the product of the amplitudes
for $\alpha_{01}$ and $\alpha_{12}$.  This rule is closely tied up with the
unitarity and linearity of quantum mechanics -- i.e., with the probability
interpretation and the principle of superposition 
-- and it would certainly 
be very difficult to get along without it.  The rule is automatically
obeyed by the usual expression for the amplitude as 
the exponential of $i$ times the classical action.

So let us determine the disconnected pieces, into which the space of 
identical particle
trajectories falls.  We need 
consider only closed trajectories,
that is trajectories with identical initial and final configurations, since these are what appear in
inner products.  
To begin with, let us focus on just two particles.

In {\it two} spatial dimensions, but not in any higher number, 
we can unambiguously define
the angle through which one particle moves with respect to the other, as
they go through the trajectory.  It will be a multiple of $\pi$; an odd
multiple if the particles are interchanged, an even multiple if they are
not.  Clearly the angle adds, if we follow one trajectory by another.  Thus
a weighting of the trajectories, consistent with the basic rule 
stated in the preceding paragraph, is
\begin{equation}
\rho(\alpha ) = e^{i\theta \phi /\pi },
\end{equation}
where $\phi$ is the winding angle, and $\theta$ is a new parameter.  As
defined, $\theta$ is periodic modulo $2\pi$.  
In three or more dimensions,
the change in the angle $\phi$ cannot be defined unambiguously. In these
higher dimensions it
is 
only defined modulo $2\pi$. 
In three or more dimensions, then, we must have
$e^{i\theta \phi /\pi} = e^{i\theta \phi ' /\pi}$ if $\phi$ and $\phi '$
differ by a multiple of $2\pi$.  So in three or more dimensions 
we are essentially reduced to the two
cases $\theta = 0$ and $\theta = \pi$, which give a factor 
of unity or a minus sign 
respectively for trajectories with interchange.  
Thus in three dimensions the preceding arguments just reproduce the familiar
cases -- bosons and fermions -- of quantum statistics, 
and demonstrate that they exhaust the
possibilities.  

In two space dimensions, however, we 
see that there are
additional possibilities for the weighting of identical particle paths. 
Particles carrying the new
forms of quantum statistics, are called generically {\it anyons}.  

Passing to $N$ particles, we find that in three or more dimensions the
disconnected pieces of trajectory space are still classified by
permutations.  With the obvious natural rule for composing paths
(as used in our statement of the consistency requirement for quantum
mechanics, above), we find that 
the disconnected pieces of trajectory space correspond to 
elements of the permutation group
$P_n$.  Thus the consistency rule, for three or more dimensions, requires
that the weights assigned amplitudes from different disconnected classes
must be selected from some representation of the group $P_n$.

In two dimensions there is a much richer
classification, involving the so-called braid group $B_n$.  
The braid group is a
very important mathematical object. 
The elements of the braid
group are the disconnected pieces of trajectory space.  
The multiplication 
law, which makes it a group,
is simply to follow one trajectory from the first piece, by another from
the second piece -- their composition lands in a uniquely determined
piece of trajectory space, which defines the group product.  The ``braid''
in braid group evidently refers to the interpretation of the disconnected
pieces of trajectory space as topologically
distinct methods of styling coils of hair.   

It may be shown that the 
braid group for $n$ particles is generated by $n-1$ generators $\sigma_k$
satisfying the relations
\begin{align}
\sigma_j \sigma_k &= \sigma_k \sigma_j\ , \qquad\qquad |j-k|  \geq 2\nonumber \\
\sigma_j \sigma_{j+1} \sigma_j  &= \sigma_{j+1} \sigma_j  \sigma_{j+1}\ ,
\quad 1 \leq j \leq n-2\ . 
\end{align}
The $\sigma$s generate counterclockwise 
permutations of adjacent particles (with
respect to some fixed ordering). 
Thus in formulating the quantum mechanics of identical particles, 
we are led to consider representations of $P_n$ -- or, in two spatial
dimensions, $B_n$.  The simplest representations are the
one-dimensional ones.  These are anyons with parameter $\theta$, as previously defined.  Higher-dimensional representations
correspond to particles with some sort of internal degree of freedom, intimately 
associated with their quantum
statistics.

This discussion of fractional statistics has been at the level of quantum particle kinematics.  Their implementation in quantum field theory
uses the so-called Chern-Simons construction.   This was spelled out for the first time in the accompanying paper of Arovas, Schrieffer, Wilczek and
Zee \cite{ASWZ}.

\section{Fractional Quantum Numbers\\ with Abstract Vortices}

For reasons mentioned before, two-dimensional systems provide an especially fertile source of fractionalization phenomenon.  In
this section I'll discuss an idealized model that exhibits the salient phenomena in stripped-down form.

Consider a 
$U(1)$ gauge theory spontaneously broken to a discrete $Z_n$ subgroup.  
In other words, we imagine that some charge $ne$ field $\phi$ condenses, and
that there are
additional unit charge particles, produced by a field $\psi$, 
in the theory.  The case $ n = 2$ is realized in ordinary BCS
superconductors, where the doubly charged 
Cooper pair field condenses, and there are
additional singly charged fields to describe the normal electron (pair-breaking) 
excitations. 
 
Such a theory supports vortex solutions \cite{A, NO}, where the $\phi$ field
behaves asymptotically as a function of the angle $\theta$ as 
\begin{equation}
\phi (r,\theta) \rightarrow  ve^{i\theta},\  r \rightarrow \infty 
\end{equation}
where $v$ is the value of $\phi$ in the homogeneous ground state.
To go with this asymptotics for $\phi$ we must have for 
the gauge potential
\begin{equation}
A_\theta(r, \theta) \rightarrow \frac{1}{ne} 
\end{equation}
in order that the covariant derivative $D_\theta\phi = (\partial_\theta
\phi- i neA_\theta)$, which appears (squared) in the energy density will 
vanish at infinity.   Otherwise the energy would diverge.  
 
In this set-up the magnetic field strength $B = \nabla \times A $ vanishes
asymptotically.  Indeed, 
since the fields transform as
\begin{align} 
\phi' (x)  &= \exp (iQ\Lambda (x)) \phi (x)  = \exp (ine\Lambda (x)) \phi
(x) \nonumber\\
A_\mu' (x) &= A_{\mu} (x) + \partial_\mu \Lambda (x) 
\end{align}
under a gauge transformation we can, by choosing $\Lambda = -\theta/ne$, remove the space
dependence of $\phi$ and make $A_{\theta}$  
vanish altogether.  We have, it appears, transformed back to the homogeneous
ground state. 
However 
$\Lambda$ is not quite a kosher gauge transformation, because the angle
$\theta$ is not a legitimate, single-valued function.  

The correct formulation
is that the vortex asymptotics is trivial, and can be gauged away, locally 
but {\it not\/} globally.  Since we can pick a well defined branch of 
$\theta$ in any
patch that does not surround the origin, all local gauge invariant
quantities must reduce to their ground state values (this explains why $D\phi$ and $F$ vanish).  
But the line integral 
$\oint A\cdot dl$ of $A$ around
a closed loop surrounding the origin, which by Stokes' theorem measures
the flux inside, cannot be changed by any legitimate gauge transformation.
And it is definitely {\it not\/} zero for the vortex; indeed we have the enclosed
flux $\Phi = \oint A\cdot dl = \frac {2\pi}{ne} $.

Another aspect of the
global non-triviality of the vortex, is that our putative gauge
transformation $\Lambda = -\theta/ne$ transforms a unit charge field
$\psi$ into something that is not single-valued.    Since
\begin{equation}
\psi'(x) = \exp (i \Lambda (x) ) \psi (x)
\end{equation}
we deduce
\begin{equation}
\psi' (\theta + 2\pi) = \exp \Bigl(-\frac{2\pi}{n}\Bigr) \psi' (\theta). 
\end{equation}
 
Now let us discuss angular momentum.  Superficially,
vortex asymptotics of the scalar order parameter seems to trash
rotational invariance.  For a scalar field should be unchanged by a rotation, but
$ve^{i\theta}$ acquires a phase.  
However we must remember that the phase of $\phi$ is
gauge dependent, so 
we can't infer from this that any {\it physical\/}
property of the vortex violates rotation symmetry.  Indeed, it is easy to
verify that if we supplement the naive rotation generator $J_z$ with an
appropriate gauge transformation
\begin{equation}
K_z = J_z - \frac{Q}{ne}
\end{equation}
then $K_z$ leaves both the action, 
and the asymptotic scalar field configuration
of the vortex invariant.  
 
Thus, assuming that the core is invariant, $K_z$ generates
a true rotation symmetry of the vortex.  If the core is
not invariant, the solution will 
have a finite moment of inertia, and upon
proper quantization we will get a spectrum of rotational excitations of the
vortex, similar to the band
spectrum of an asymmetric molecule.  This step, of course, does not introduce any fractional angular momentum.  

For present purposes, the central point is that 
passing from $J$ to $K$ modifies the quantization condition 
for angular momentum
of quanta orbiting the vortex.  
In general, their orbital angular momentum becomes fractional.
The angular momentum of
quanta with the fundamental charge $e$, for example, is quantized in units
of $-\frac{1}{n} + {\rm integer}$.  

In two space dimensions the object consisting of a vortex together with its orbiting electron will appear as a
particle, , since its energy-momentum distribution is well localized.  But it carries a topological charge, of boundary type.   
That is
the secret of its fractional angular momentum.

The general connection between spin and statistics suggests that objects with fractional angular
momentum should likewise carry fractional statistics.  Indeed there is a very general argument, the
ribbon argument of Finkelstein and Rubenstein \cite{FR}, which connects particle interchange and
particle rotation.    The space-time process of creating two particle-antiparticle pairs, interchanging the
two particles, and finally annihilating the re-arranged pairs, can be continuously deformed into the
process of creating a pair, rotating the particle by $2\pi$, and finally annihilating.   Therefore, in a path
integral, these two processes must  be accompanied by the same non-classical phase.   This leads to 
\begin{equation}
P = e^{2\pi i S},
\end{equation}
where $S$ is the spin and $P$ is the phase accompanying (properly oriented) interchange.  This gives the ordinary spin-statistics
connection in 3+1 space-time dimensions, in a form that generalizes to anyons.   For our vortex-$\psi$ composites, it is easy to
see how the funny phase arises.   It is a manifestation of the Aharonov-Bohm effect \cite{AB}.  Transporting charge $e$ around flux
$1/ne$  -- or, for interchange, half-transporting two such charges around one anothers'  fluxes -- accumulates
non-classical phase $2\pi/n$.

\section{Fractional Quantum Numbers\\ in the Quantum Hall Effect}

Microscopic understanding of the fractional quantum Hall effect has been built up from Laughlin's variational wave function, analogously to how
microscopic understanding of superconductivity was built up from the BCS variational wave function \cite{L, PG}.   To be concrete, let us consider the
$\frac{1}{3}$ state.  The ground state wave function for $N$ electrons in a droplet takes the form
\begin{equation}
\psi (z_1, ... , z_N) ~=~ \prod_{i<j} (z_i - z_j)^3 \prod _i \exp (-|z_i|^2/l^2)
\end{equation}
where $l^2 \equiv \frac{4}{eB}$ defines the magnetic length, and we work in symmetric gauge $A_x = -\frac{1}{2} By, A_y = +\frac{1}{2} Bx$.   

The
most characteristic feature of this wave function is its first factor, which encodes electron correlations.  Through it, each electron repels other
electrons, in a very specific (holomorphic) way that allows the wave function to stay entirely within the lowest Landau level.  
Specifically, if electron 1
is near the origin, so
$z_1 = 0$, then the first factor contributes
$\prod_{1<i} z_i^3$.  This represents, for each electron, a boost of three units in its angular momentum around the origin.   (Note that in the
lowest Landau level the angular momentum around the origin is always positive.)

Such a universal kick in angular momentum has a simple physical interpretation, as follows.  
Consider a particle of charge $q$
orbiting around a thin solenoid located 
along the
$\hat z$ axis.  Its angular momentum along the $\hat z$ axis evolves
according to 
\begin{align}
\frac{dL}{dt} &= qrE_\phi \nonumber\\
             &= \frac{q}{2\pi} \frac{d\Phi}{dt}
\end{align}
where $E_\phi$ is the value of the azimuthal electric field and
$\Phi$ is the value of the flux through the solenoid; the second equation
is simply Faraday's law.  Integrating 
this simple equation we deduce the simple
but profound conclusion that
\begin{equation}
\Delta L = \frac{1}{2\pi} \Delta (q\Phi ) .
\end{equation}
The change in angular momentum is equal to the change in the 
flux times charge.  All details about how the flux got built up cancel out. 

From this point of view, we see that in the $\frac{1}{3}$ state each electron implements correlations as if it were a flux tube with flux $3
\frac{2\pi}{q}$.  This is three times the minimal flux.   Now let us follow Schrieffer's idea, as previously discussed for polyacetylene, and remove the
electron.  This produces a hole-like defect, but one that evidently, as in polyacetylene, begs to be broken into more elementary pieces.   
Either from the flux point of view, or directly from the wave function, it makes sense to break consider an elementary quasi-hole of the type
\begin{equation}
\psi (z_2, ... , z_N) ~=~ \prod z_i \prod_{i<j} (z_i - z_j)^3 \prod _i \exp (-|z_i|^2/l^2) . 
\end{equation}
 (Note that electron 1 has been removed.)  The first factor represents the defect.  By adding three defects and an electron, we get back to the ground
state.   Thus the elementary quasihole will carry charge $-e/3$.  

Here again the Schrieffer counting argument is correct and utterly convincing, but a microscopic derivation adds additional insight. 
It is given in the accompanying paper by Arovas, Schrieffer, and Wilczek \cite{ASW}, through an orchestration of Berry's phase and the Cauchy
integral theorem.  

At another level of abstraction, we can use the Chern-Simons construction to model the electrons as being vortices, quite literally, 
of a fictitious gauge
field.  This leads to a profound insight into the nature of the quantum Hall effect, which ties together most of what we've discussed, and provides an
appropriate climax.   

A constant magnetic field frustrates condensation of electrically charged particles, because the gradient energy 
\begin{equation}
\int |\partial \eta - i  q_{\rm el.} A e \eta |^2 \sim (qe)^2 |\langle \eta \rangle|^2  \int A^2 
\end{equation}
grows faster than the volume, due to the growth of $A$, and therefore cannot be sustained.  This is the theoretical root of the Meissner effect. 
However if each particle acts as a source of fictitious charge and flux, then
the long-range part of the total potential $q_{\rm el.} e A + q_{\rm fict.} a$ will vanish, and the frustration will be removed, if
\begin{equation}
q_{\rm el.} e B + q_{\rm fict.} n_\eta \Phi_{\rm fict.} ~=~ 0, 
\end{equation}
where $n_\eta$ is the number-density of $\eta$ quanta and $\Phi_{\rm fict.}$ is the fictitious flux each
carries.  Given $\frac{q_{\rm el.}eB}{n}$ -- that is to say, a definite filling fraction -- a definite value of
$q_{\rm fict.} \Phi_{\rm fict.}$ is implied.   But it is just this parameter that specifies how the
effective quantum statistics of the $\eta$ quanta have been altered by their fictitious gauge charge and
flux.   Condensation will be possible if -- and only if -- the altered statistics is bosonic.   Identifying the
$\eta$ quanta as electrons, we require  
\begin{equation}
q_{\rm fict.} \Phi_{\rm fict.}  =  (2m+1) \pi 
\end{equation}
with $m$ integral, to cancel the fermi statistics.  We also have $\frac{q_{\rm el.}eB}{n} = \frac{eB}{n} = \frac{\pi}{\nu}$, for filling fraction $\nu$. 
Thus we derive  
\begin{equation}
\frac{1}{\nu} = 2m +1, 
\end{equation}
accounting for the primary Laughlin states.   

These connections among superconductivity, statistical transmutation, and the quantum Hall effect can be extended conceptually, to bring in anyon
superconductivity \cite{lFHL, CWWH} and composite fermions \cite{J}; tightened into what I believe is a physically rigorous derivation of the quantum Hall
complex, using adiabatic flux trading \cite{W}; and generalized to multi-component systems (to describe multilayers, or states where both directions of spin
play a role) \cite{WZ}, and more complicated orderings, with condensation of pairs \cite{MR, GWW}.   In this field, as in many others, the fertility of Bob
Schrieffer's ideas has been invigorated, rather than exhausted, with the harvesting.  

\subsubsection*{Acknowledgments}
This work is supported in part by
funds provided by the U.S. Department of Energy (D.O.E.) under cooperative research
agreement
No.\ DF-FC02-94ER40818.   I would like to thank Reinhold Bertlmann and the University of Vienna,
where this work was completed, for their hospitality.


\begin{thebibliography}{99}

\bibitem{JR} R. Jackiw and C. Rebbi, Phys. Rev. {\bf D13}, 3398 (1976).

\bibitem{SSH} W. P. Su, J. R. Schrieffer, and A. J. Heeger, Phys. Rev. Lett. {\bf 42}, 1698 (1979); Phys. Rev. {\bf B22}, 2099 (1980). 

\bibitem{SS} W. P. Su and J.R. Schrieffer, Phys. Rev. Lett. {\bf 46}, 738 (1981).

\bibitem{JS} R. Jackiw and J. R. Schrieffer, Nucl. Phys. {\bf B190}, 253 (1981).

\bibitem{P} R. F. Peierls, {\it Quantum Theory of Solids\/} (Clarendon Press, Oxford, 1955).

\bibitem{CW} S. Coleman and E. J. Weinberg, Phys. Rev. {\bf D7}, 1888 (1973).

\bibitem{K} D. B. Kaplan, Phys. Lett. {\bf B301}, 219 (1993).

\bibitem{HFSW} T. L. Ho, J. R. Fulco, J. R. Schrieffer, and F. Wilczek, Phys. Rev. Lett. {\bf 52}, 1524 (1984).

\bibitem{GW} J. Goldstone and F. Wilczek, Phys. Rev. Lett. {\bf 47}, 986 (1981).

\bibitem{GG} H. Georgi and S. Glashow, Phys. Rev. Lett. {\bf 32}, 438 (1974); see also J. Pati and A. Salam, Phys. Rev. {\bf  D8}, 1240 (1973).

\bibitem{D} P. A. M. Dirac, Proc. R. S. London, {\bf A133}, 60 (1931).

\bibitem{'t} G. 'tHooft, Nucl. Phys. {\bf B79}, 276 (1974).

\bibitem{Po} A. Polyakov, JETP Lett. {\bf 20}, 194 (1974).

\bibitem{S} T. H. R. Skyrme, Proc. R. S. London, {\bf 260}, 127 (1961).

\bibitem{R} Reviewed in R. Rajaraman, {\it Solitons and Instantons}, (North Holland, 1982). 


\bibitem{LM} J. M. Leinaas and J. Myrheim, Il Nuovo Cimento {\bf 50}, 1 (1977); F. Wilczek, Phys. Rev. Lett. {\bf 48}, 1144 (1982); G. A. Goldin, R.
Menikoff, and D. H. Sharp, Phys. Rev. Lett. {\bf 51}, 2246 (1983).

\bibitem{ASWZ} A. Arovas, J. R. Schrieffer, F. Wilczek, and A. Zee, Nucl. Phys. {\bf B251}, 917 (1985).

\bibitem{A} A. A. Abrikosov, Zh. Eksp. Teor. Fiz. {\bf 32}, 1442 (1957).

\bibitem{NO} H. B. Nielsen and P. Olesen, Nucl. Phys. {\bf B61} 45 (1973).

\bibitem{FR} D. Finkelstein and J. Rubinstein, J. Math. Phys. {\bf 9}, 1762 (1968).

\bibitem{AB} Y. Aharonov and D. Bohm, Phys. Rev. {\bf 115}, 485 (1959).

\bibitem{L} R. B. Laughlin, Phys. Rev. Lett. {\bf 50}, 1395 (1983).

\bibitem{PG}  Reviewed in R. Prange and S. Girvin, {\it The Quantum Hall Effect}, (Springer Verlag, 1990).

\bibitem{ASW} D. Arovas, J. R. Schrieffer, and F. Wilczek, Phys. Rev. Lett. {\bf 53}, 2111 (1984).

\bibitem{lFHL} R. B. Laughlin, Science {\bf 242}, 525 (1988); A. L. Fetter, C. B. Hanna, and R. B. Laughlin, Phys. Rev. {\bf B39}, 9679 (1989).

\bibitem{CWWH} Y.-H. Chen, F. Wilczek, E. Witten, and B. I. Halperin, Int. J. Mod. Phys. {\bf B3}, 903 (1989).

\bibitem{J} J. K. Jain, Phys. Rev. Lett. {\bf 63}, 199 (1989).

\bibitem{HLR}  B. I. Halperin, P. A. Lee, and N. Read, Phys. Rev. {\bf B47}, 7312 (1993).


\bibitem{W} F. Wilczek, Int. J. Mod. Phys. {\bf B5}, 1273 (1991); M. Greiter and F. Wilczek, Mod. Phys. Lett. {\bf B4}, 1063 (1990).

\bibitem{WZ} X. G. Wen and A. Zee, Phys. Rev.  {\bf B46}, 2290 (1992).

\bibitem{MR} G. Moore and N. Read, Nuc. Phys. {\bf B360}, 362 (1991).

\bibitem{GWW} M. Greiter, X. G. Wen, and F. Wilczek, Phys. Rev. Lett. {\bf 66}, 3205 (1991).
 

\end{thebibliography}
\end{document}